\pdfoutput=1
\documentclass[graybox]{svmult}

\usepackage{mathptmx}       
\usepackage{helvet}         
\usepackage{courier}        
\usepackage{type1cm}        
\usepackage{makeidx}         
\usepackage{graphicx}        
\usepackage{multicol}        
\usepackage[bottom]{footmisc}
\usepackage{xcolor}

\usepackage[nolist,nohyperlinks]{acronym}
\begin{acronym}[ICANN]
    \acro  {fl}   [FL]   {Federated Learning}
    \acro  {oop}   [OOP]   {Object Oriented Programming}
    \acro  {ml}   [ML]   {Machine Learning}
    \acro  {ds}   [DS]   {Distributed Systems}
    \acro  {ai}   [AI]   {Artificial Intelligence}
    \acro  {cfl}   [CFL]   {Centralized Federated Learning}
    \acro  {sdfl}   [SDFL]   {Semi-decentralized Federated Learning}
    \acro  {dfl}   [DFL]   {Decentralized Federated Learning}
    \acro  {iot}   [IoT]   {Internet-of-Things}
    \acro  {dai}   [DAI]   {Distributed Artificial Intelligence}
    \acro  {flops} [FLOPS] {Floating point operations per second}
    \acro  {hpc} [HPC] {High-Performance Computing}
    \acrodefplural{iot}[IoTs]{Internets-of-Things}
\end{acronym}

\usepackage{hyperref}

\makeindex             

\begin{document}

\title*{Falafels: A tool for Estimating Federated Learning Energy Consumption via Discrete Simulation}
\titlerunning{Falafels: Estimating Federated Learning Energy Consumption}
\author{Andrew Mary Huet de Barochez, Stéphan Plassart and Sébastien Monnet}
\institute{
    Andrew Mary Huet de Barochez 
    \and Stéphan Plassart
    \and Sébastien Monnet
    \at LISTIC, Université Savoie Mont-Blanc, Annecy, France 
    \email{{andrew.mary-huet-de-barochez, stephan.plassart, sebastien.monnet}@univ-smb.fr}
}

%
%
\maketitle

\abstract{
The growth in computational power and data hungriness of Machine Learning has led to an important shift of research efforts towards the distribution of ML models on multiple machines, leading in even more powerful models.
However, there exists many Distributed Artificial Intelligence paradigms and for each of them the platform and algorithm configurations play an important role in terms of training time and energy consumption.
Many mathematical models and frameworks can respectively predict and benchmark this energy consumption, nonetheless, the former lacks of realism and extensibility while the latter suffers high run-times and actual power consumption.
In this article, we introduce \texttt{Falafels}, an extensible tool that predicts the energy consumption and training time of — but not limited to — Federated Learning systems.
It distinguishes itself with its discrete-simulator-based solution leading to nearly instant run-time and fast development of new algorithms.
Furthermore, we show this approach permits the use of an evolutionary algorithm providing the ability to optimize the system configuration for a given machine learning workload.
}


\section{Introduction}
\label{sec:introduction}

\sloppy 

Since 2012, \ac{ai} computing power is growing exponentially~\cite{OpenaiRef}. 
As a consequence, \acf{dai}~\cite{JANBI2023200231}, i.e., combining the computing power of multiple machines, is imperative.
However, this exponential growth considerably worsen the impact of \ac{ai}~\cite{patterson2021carbon} on our planet's limited resources~\cite{richardson2023earth}, therefore evaluating and reducing energy consumption of such systems is critical.
To account for this consumption, we distinguish two complementary approaches:
\begin{itemize}
\item \textbf{Measuring} the electricity for the training, either with software-based solutions or with a watt-meter.
\item \textbf{Estimating} the consumption via an energy model which takes as an input the execution time. This last can be retrieved a posteriori, or it can be predicted a priori by modelling the execution order.
\end{itemize}

In classical \ac{ai}, studying and optimizing the model is the main way of reducing energy consumption~\cite{sze2017efficient, yang2017designing}, but it is now insufficient in a context where the training workload is parallelized on multiple machines.
Indeed, now communication cost must be taken into account, the platform on which the model will be trained must be sized and configured properly, and the algorithms used for machine coordination must be tested extensively as they play an important role on training time and energy consumption \cite{verbraeken2020survey}.
As a consequence, many parameters related to the system configuration exist and need to be chosen correctly.

A particularly interesting case of application is the \acf{fl} paradigm~\cite{mcmahan2017communication} which leverages the computational power of several heterogeneous devices with limited computational power, bandwidth, memory storage, and availability.
This aspect makes energy measurement and estimation more complex because the technique employed has to be adapted to each device, compared to an homogeneous cluster with the same machines.
Furthermore, one can configure for example the applicative topology, the communication mechanism, the decentralization level, or the scheduling algorithm.
This creates many combinations of parameters, making it difficult for measuring-based technologies as it requires to re-run the training for every combination, representing actual energy consumption and long run-times.

In this paper, we present \texttt{Falafels}\footnote{\url{https://github.com/PhoqueEberlue/falafels}}, a tool specifically built to study the \ac{ds} challenges of \ac{dai}.
\texttt{Falafels} provides extensible building blocks for simulating and estimating the energy consumption of a \ac{fl} system using a single-node.
It is based on discrete simulation using the \texttt{Simgrid} framework\footnote{\url{https://simgrid.org/}}, which abstracts the computation workloads to focus on networking, task ordering and scheduling.
Furthermore, \texttt{Simgrid} provides an energy model~\cite{heinrich2017predicting} that is used in \texttt{Falafels} in order to estimate the energy consumed by training machines and network devices.
We believe that discrete-simulation-based tools are complementary to experimental frameworks because they have faster execution times, providing quick evaluations of distributed algorithms.
The following contributions of this paper are:
\begin{enumerate}
    \item We propose an extensible modeling of \ac{fl} using the object oriented programming paradigm, providing seamless interoperability between networking and learning algorithms.
    \item We introduce \texttt{Falafels}, a tool for simulating \ac{fl} systems execution and predicting energy consumption of the machines and the network with a nearly instant run-time.
        This provides the ability to run fast experimentations of many configuration parameters and to develop new distributed algorithms for parallelizing efficiently machine learning tasks.
    \item We demonstrate how optimization algorithms such as an evolutionary algorithm can be applied using our tool in order to find an optimal system configuration parameters for a given workload.
\end{enumerate}

The remainder of this article is organized as follows. 
Sec.~\ref{sec:related-work} details the principles and parameters of \ac{fl} systems and provides references to related works concerning energy optimization in distributed systems.
Sec.~\ref{sec:falafels} introduces the developed tools and the state-of-the-art algorithms that have been implemented.
Sec.~\ref{sec:evolution} details the evolution algorithm and experimental setup used before Sec.~\ref{sec:conclusion} concludes the paper and gives some perspectives.

\section{Background and Related work}
\label{sec:related-work}

\acf{fl} is a trending domain introduced by Google~\cite{mcmahan2017communication}, and its particularity of using many heterogeneous devices makes it challenging to find the best configurations for a use case.
Due to the network topologies and algorithms available, \cite{lo2022architectural} introduce a large number of different configurations. \cite{beltran2023decentralized} separates \ac{fl} systems in three categories.

The \textbf{Federation Architecture} includes the \textit{federation type}; Cross-silo refers to collaborations between organizations owning large datasets and powerful machines, e.g, a cohort of hospitals~\cite{xu2021federated}; Cross-device leverages the power of hundred or more \ac{iot} devices \cite{savazzi2020federated}, however with limited and heterogeneous resources. 
\textit{Decentralization scheme}; \ac{cfl} is the classical scheme where a central server receives and aggregates the models; \ac{sdfl} uses edge servers to perform pre-aggregations in the middle layers of the network \cite{briggs2020federated}; \ac{dfl} can use any topology and often uses dynamic connections and role distribution~\cite{roy2019braintorrent}.

\textbf{Network topology} refers to the degree of connection between the nodes. 
\textit{Fully connected} networks provides highly reliable and robust networks as multiple points of connection between nodes prevent single-point-of-failures, however their scalability and flexibility are very low.
\textit{Partially connected} includes stars, rings, trees and random networks.
They provide better scalability and maintainability but they might be prone to single-point-of-failures.
\textit{Node clustering} generally applies to SDFL (with edge and central servers) or DFL (with less organized clusters and no central server compared to SDFL).
The advantages of this topology heavily depend on subclusters' topologies.

\textbf{Communication mechanism} describes the way nodes communicate between each others. For example, it can be the protocol (TCP/UDP/QUIC/WebSocket) or higher level concepts (P2p/Gossip/Rendez-vous). 
It also describes the level of synchronization between the rounds, for example in~\cite{chen2020asynchronous, xie2019asynchronous}, algorithms with asynchronous rounds are implemented and lead to less idle clients with faster convergence speed of the model.

\hfill

Accounting energy consumption is done either via \textbf{estimating} or \textbf{measuring}.
\cite{savazzi2022energy, pilla2023scheduling} implemented mathematical frameworks in order to \textbf{estimate} \ac{fl} systems consumption, they found out that communication efficiency and trainer population size are important parameters when optimizing energy consumption.
Preliminary work have also been done concerning the carbon footprint of \ac{fl}~\cite{qiu2023first}.
However mathematical frameworks have limitations concerning the realness of the results because of the lack of system considerations.

Many experimental frameworks exist~\cite{bonawitz2019towards, lai2022fedscale, caldas2018leaf}, and energy consumption can be \textbf{measured} using benchmarking softwares or watt-meters.
Nevertheless, the previous frameworks tend to be limited in realism and scalability, particularly when the \ac{fl} system is simulated on a single-node.
\texttt{Flower}~\cite{beutel2020flower} and \texttt{ns3-fl}~\cite{ekaireb2022ns3} are two frameworks built to reproduce systems limitations in order to provide closer results between experimentation and production.
The former proposes a flexible application where experiments can be executed both on a single-node and in real conditions on multiple nodes, and this without modifying the code.
The latter paired the existing framework \texttt{flsim}~\cite{wang2020optimizing} with the discrete network simulator \texttt{ns-3}~\cite{riley2010ns} in order to simulate the \ac{fl} system and the network on a single-node.

Mathematical frameworks lack realism and extensibility while experimental frameworks are constrained by the \ac{ml} model training time, thus, researching and evaluating new solutions for the \ac{ds} challenges, e.g., implementing a new fault-tolerance algorithm, can become tedious.
From \ac{ds} point of view, we want to abstract the \ac{ml} workload and focus on the distributed algorithms themselves.
To answer this problematic, we pushed the idea of \texttt{ns3-fl} where only the network is discrete, to a fully discrete simulation framework called \texttt{Falafels}.
This led to nearly instant execution times, energy consumption is estimated which reduces the impact of experimenting, and the experimentation of new distributed algorithms becomes easier.
This comes at the cost of the inability to know about \ac{ml} model accuracy, making it a complementary approach to existing frameworks.
\texttt{Falafels} currently only implements \ac{fl} systems simulation, but the architecture is flexible enough to embrace others paradigms such as Distributed Deep Learning by reducing the granularity of computation and communication tasks as performed in \texttt{ASTRA-sim2.0} \cite{won2023astra}.

\section{Falafels: federated learning with discrete simulation}
\label{sec:falafels}

In this section, we introduce the use cases, design goals, a formalization of \ac{fl} concepts and algorithms, and finally the core architecture of our framework \texttt{Falafels}.

\subsection{Use Cases}

Existing tools have mostly focused on the machine learning aspect of \ac{fl}, making them specialized in research towards model performance while neglecting systems related aspects.
The goal of \texttt{Falafels} is to provide a convenient tool for analyzing and researching new network architectures and distributed algorithms for \ac{fl} by mostly focusing on energy consumption rather than model performance.
This tool aims to be complementary with the existing frameworks in order to find compromises in terms of model performance and energy consumption.

\textbf{Energy prediction for machines and network.}
In \ac{dai} and \ac{fl} in particular, networking can consume a lot of energy. The experiments must be able to predict the energy consumption from both the computations and the communications.

\textbf{Large scale network experiments.}
The simulations need to be fast in order to conveniently iterate over results, and this statement should remain true even when working with large scale networks.

\textbf{Single-node execution.}
Simulating a whole system on a single-node is very convenient for developing and debugging new distributed algorithms, it provides access to a global vision of the network, allowing to better understand the impacts of the new solutions.

\textbf{Reproducibility of experiments.}
Providing an environment that can be reproduced by other researchers is crucial to propose quality research.
Furthermore, reproducibility within a discrete event simulator has the advantage of having a deterministic history of network messages and task execution which is practical for debugging, compared to real experiments for which behaviours can diverge from multiple executions.

\subsection{Design goals}

\texttt{Falafels} focuses on use cases concerning~\ac{ds} problems applied to~\ac{fl}. 
However we do not want to restrict the design and implementation to the~\ac{fl} case, instead it should be open to other~\ac{dai} paradigms for the future.

\textbf{Learning Algorithm extensibility.}
The simulator should be extensible, adding new algorithms or modifying existing ones must be simple with a base level of abstraction for communications.

\textbf{Network topologies plurality.}
Given the number of network topologies available in \ac{ds}, we want to provide multiple networking algorithms built on common abstract primitives so that the learning algorithms are seamlessly compatible with any networking algorithm.

\textbf{Nodes can change role at run-time.}
To permit research in dynamic and decentralized methods, nodes should be able to change their role at run-time.
This is easier to work with this requirement from the start than adding it later during development.

\subsection{Federated Learning modelling}

\begin{figure}[h]
    \centering
    \includegraphics[width=0.9\linewidth]{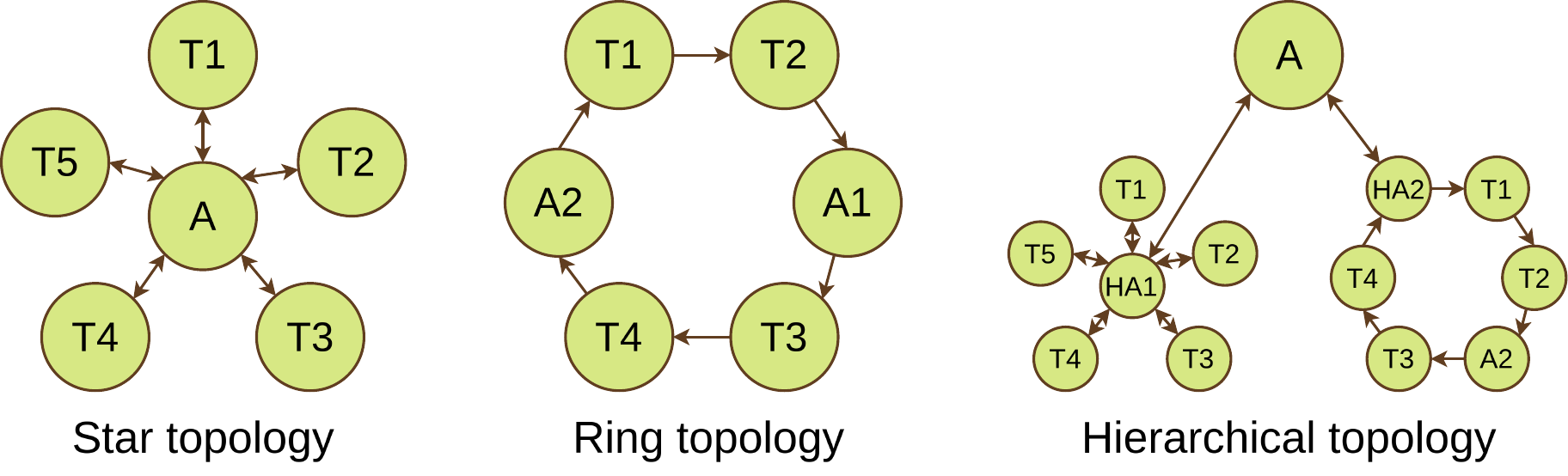}
    \caption{Implemented network topologies. In a star, each trainer T$_x$ is connected to a central aggregator that orchestrates the training. The ring is unidirectional, packets flow in only one direction and multiple aggregators A$_y$ can be used. Hierarchical topology allows the connection of multiple subclusters to the central aggregator via hierarchical aggregators HA$_z$.}
    \label{fig:topologies}
\end{figure}

Following the state-of-the-art in the~\ac{fl} domain (see Sec.~\ref{sec:related-work}), we identified the main algorithms and network topologies in order to support a broad part of existing implementations, while also providing a base for further extensions.
Fig.~\ref{fig:topologies} illustrates the different topologies implemented; the star topology is used mostly in~\ac{cfl} and it is rarely used in~\ac{dfl}; the ring topology is typically used in~\ac{dfl}; the hierarchical topology can be used both for edge computing (\ac{sdfl}) and clustering (\ac{dfl}).

\begin{figure}[h]
    \centering
    \includegraphics[width=0.8\linewidth]{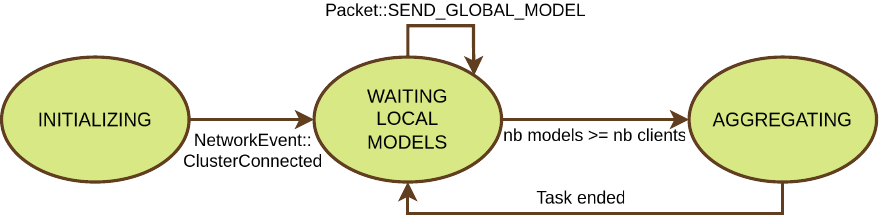}
    \caption{Automaton representing the simple aggregator algorithm.}
    \label{fig:simple-aggregator-automaton}
\end{figure}

The definition of an algorithm is done via a class which follows a finite-state machine behaviour.
For example, the simple aggregator has 3 states and clearly defined transitions that will be triggered upon receiving network events, or when internal class conditions are met (see Fig.~\ref{fig:simple-aggregator-automaton}).
For the asynchronous aggregator, the transition condition to the \textit{aggregating} state is set in order to wait for a given proportion of the trainers.

\begin{figure}[h]
    \centering
    \includegraphics[width=0.8\linewidth]{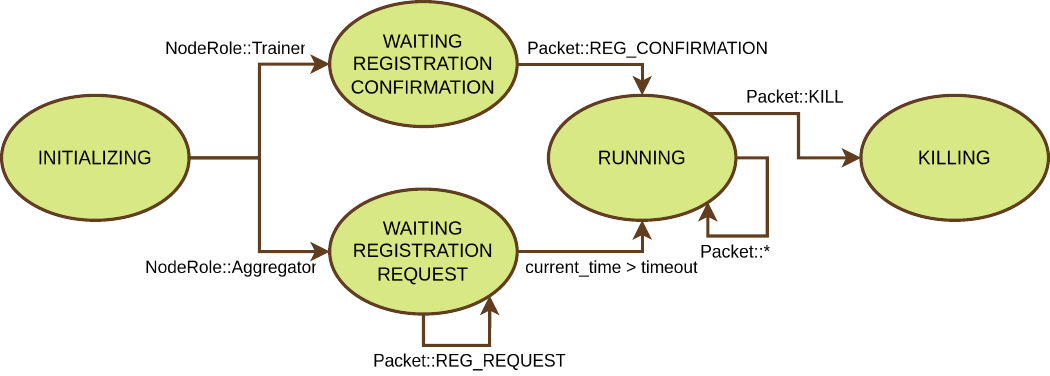}
    \caption{Example of the network manager automaton for the star topology.}
    \label{fig:networkmanager-automaton}
\end{figure}

The behaviour of networking algorithms is modelled in the same fashion.
However, we need to specify how to connect to other nodes. Depending on its role, different behaviours should be implemented, e.g., a trainer will send a registration request to the aggregator, and the aggregator will listen to those requests (see Fig.~\ref{fig:networkmanager-automaton}).
The communication logic is then executed in the \textit{running} state: if the received packet is targeted to the current node, it is sent to the role algorithm automaton, otherwise it can be redirected to another node.

\subsection{Core architecture}

As stated before, \texttt{Falafels} is based on discrete event simulation.
The two main discrete simulators renowned in the~\ac{ds} literature are \texttt{Simgrid}~\cite{heinrich2017predicting} and \texttt{ns-3}~\cite{riley2010ns}. 
Although they both have an energy model, \texttt{Simgrid} seemed better for the following reasons:

\begin{itemize}
    \item \texttt{Simgrid} is simpler, more flexible and provides higher level abstraction which is handful to focus on the application level development.
    \item \texttt{ns-3} simulate each packet in the network individually which is oversized for our use case and results in slower simulations (see \texttt{ns3-fl} run-times~\cite{ekaireb2022ns3}).
\end{itemize}

\begin{figure}[t]
    \centering
    \includegraphics[width=0.7\linewidth]{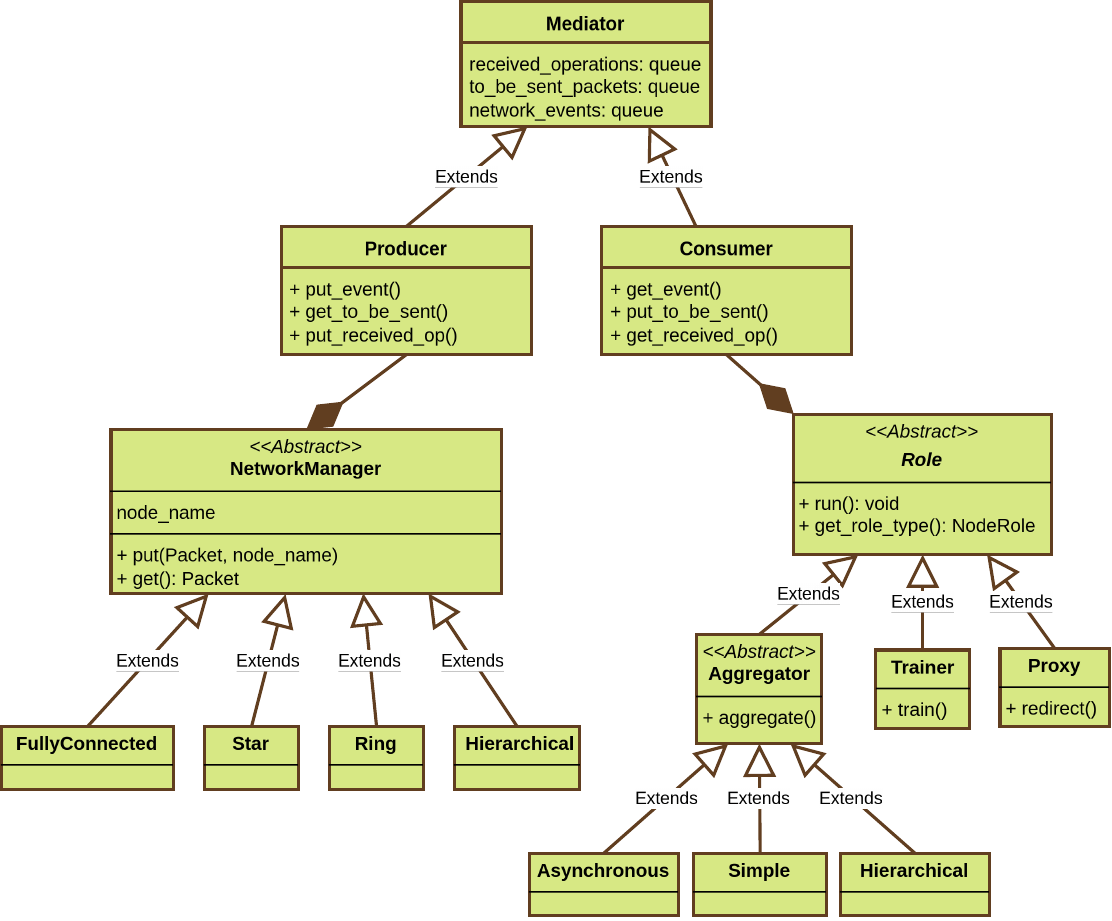}
    \caption{Class diagram of the simulator architecture.}
    \label{fig:class-diagram}
\end{figure}

The architecture is based on object oriented programming and revolves around 3~main classes: \textit{Mediator}, \textit{NetworkManager} and \textit{Role}.
Each of these classes leverages polymorphism to provide base abstraction and extendability for future features.
For example, the simplified class diagram at Fig.~\ref{fig:class-diagram} shows how a \textit{NetworkManager} provides basic \textit{put} and \textit{get} abstract functions that will be implemented differently depending on the specific implementations of \textit{NetworkManager}. 
Indeed, in a star topology, making a broadcast is as simple as sending a packet to everyone, whereas in a ring topology we want to implement a mechanism of redirection to make the packet flow around the ring.

The same applies to the \textit{Role} class, but this time there is one more layer of complexity.
Indeed, multiple layers of heritage are necessary to provide the three basic roles (Aggregator, Trainer and Proxy), and to allow heritage for their derivatives.
This makes it possible to implement multiple algorithms of Aggregators with their own execution function.
This function overrides \textit{Role}'s run function which makes each \textit{Role} children class executable in the same way, but with different behaviours.

\begin{figure}
    \centering
    \includegraphics[width=0.6\linewidth]{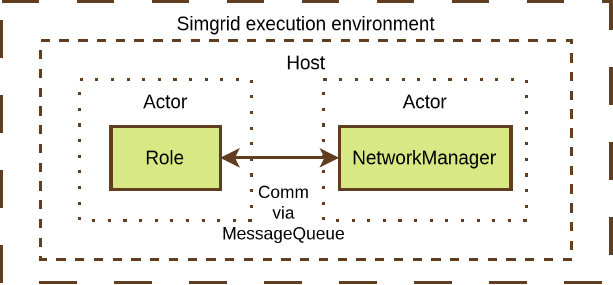}
    \caption{Execution environment of the simulator in \texttt{Simgrid}. The actual simulations contains multiples hosts (with two actors in each), however only one host have been represented here for simplicity.}
    \label{fig:execution-env}
\end{figure}

Furthermore, \textit{Role} and \textit{NetworkManager} have to be executed in parallel so that networking tasks and \ac{ml} tasks do not block each others.
\texttt{Simgrid} proposes a model with three layers: the execution environment that contains the whole simulated system, the \textbf{Host} that represent a single physical machine, and the \textbf{Actor} which corresponds to a simulated thread on a given machine (see Fig.~\ref{fig:execution-env}).
Here, \textit{NetworkManager} and \textit{Role} run on two different \textbf{Actors} while remaining on the same \textbf{Host}, so the consumption is accounted for the same machine.
This allows \texttt{Simgrid} to wait independently for events in \textit{Role} and \textit{NetworkManager}.

Finally, \textit{Mediator} allows communication between \textit{Role} and \textit{NetworkManager} as they run on two different (simulated) threads.
It is based on \textit{MessageQueues}, an equivalent of channels between threads.

\section{Evolution algorithm}
\label{sec:evolution}

\begin{figure}[h]
    \centering
    \includegraphics[width=0.8\linewidth]{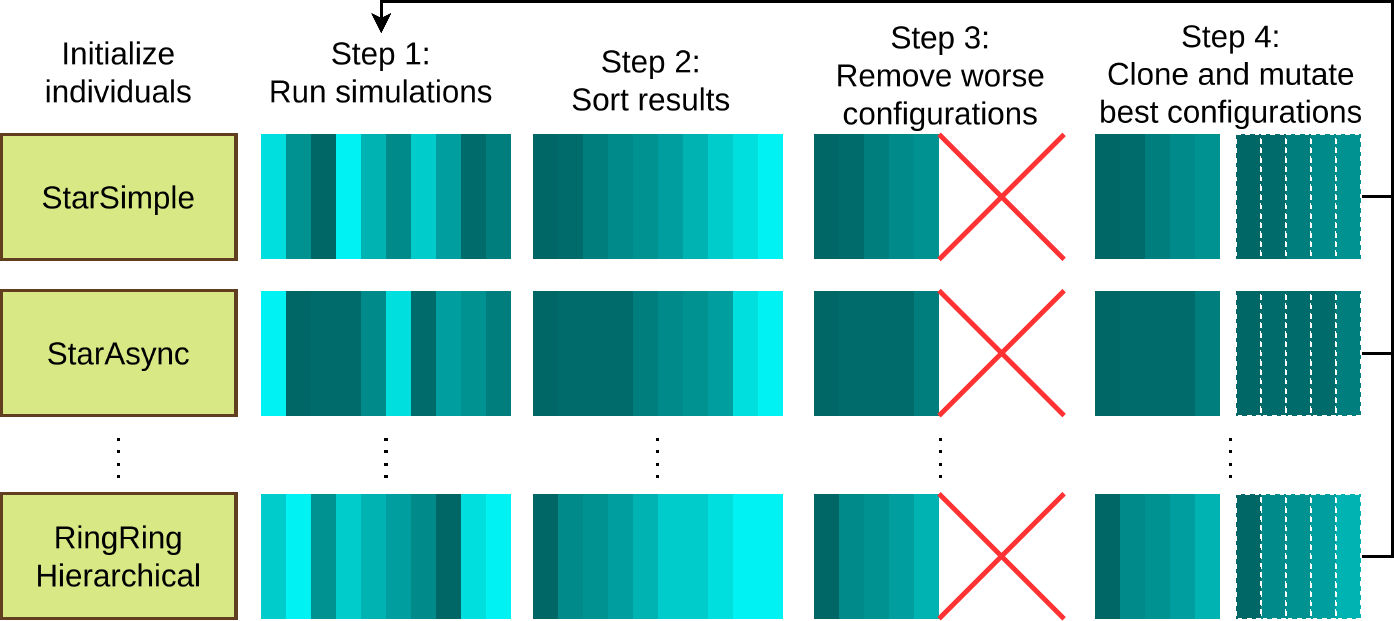}
    \caption{The evolution algorithm implemented as an example for optimizing a platform for a given \ac{ml} model. Each combination of network topology and role algorithm have its own pipeline, each pipeline contains multiple platform configuration strategies. The color gradient represent the individual performance of each strategy: the darkest are the best results.}
    \label{fig:evolution-algorithm}
\end{figure}

The evolution algorithm has been introduced to our experiment toolkit to test a large amount of simulation configurations without making any assumptions concerning their parameters. 
It can find optimal configurations by testing random parameters, thus converging as bad configurations will be filtered at each iteration based on a given criteria.
The algorithm first initializes a group for each possible combination of \textit{NetworkManager} and \textit{Role}. In each of these groups, a given number of~\ac{fl} configurations will be created. Then the following pipeline gets executed for each group (see Fig.~\ref{fig:evolution-algorithm}):
\begin{itemize}
    \item Step 1: run the simulation for each~\ac{fl} configuration of the group;
    \item Step 2: sort the results by some criteria, e.g., simulation time or total energy consumption;
    \item Step 3: remove a proportion of the worse individuals;
    \item Step 4: clone the remaining individuals and apply mutations.
\end{itemize}
Two important choices have been made while designing the algorithm.
Firstly, we separated the combinations into their respective pipeline.
It was not the case originally: the combinations with great configurations from the start would take over and prevent the other combinations to converge. It made more sense to separate them so that they would all have a possibility to converge.

Secondly, we implemented mutation on single individuals, because using reproduction between two configurations would have been harder to implement. Indeed, reproduction between two platforms does not look intuitive. Instead we used mutations, so the algorithm randomly increases (or decreases) the number of machines, changes algorithms parameters, and swaps machines roles.
To illustrate the last point, one aggregator may be running on a powerful machine during one iteration, and at the next one it could be affected to another machine with a slower profile.
\subsection*{Framework evaluation}
%
\textbf{Experimental setup:} For reproducibility purpose, all parameters and execution scripts are available online \footnote{\url{https://github.com/PhoqueEberlue/falafels/tree/740ac67ec300b8446693cfe0086491e3476b3abc/experiments/evolution/total_consumption_criteria}}.
Furthermore, many other experiments have been performed to compare the impact of platform configuration and are available on the online repository.
For the workload, we choose the model used in the first~\ac{fl} paper~\cite{mcmahan2017communication}, a multilayer-perceptron with $199,210$ total parameters.
The training task is expressed in \ac{flops} and is obtained by multiplying the number of parameters, the number of floating point operations, and the number of samples.
The same applies for the aggregation task where a weighted arithmetic mean is applied with each trainer model. Concerning the platform, we used (have simulate) three types of computers: a workstation, a laptop, and a raspberry pi~$4$. We used a benchmarking software to determine the energy model of each machines.

\begin{figure}[h]
    \centering
    \includegraphics[width=1\linewidth]{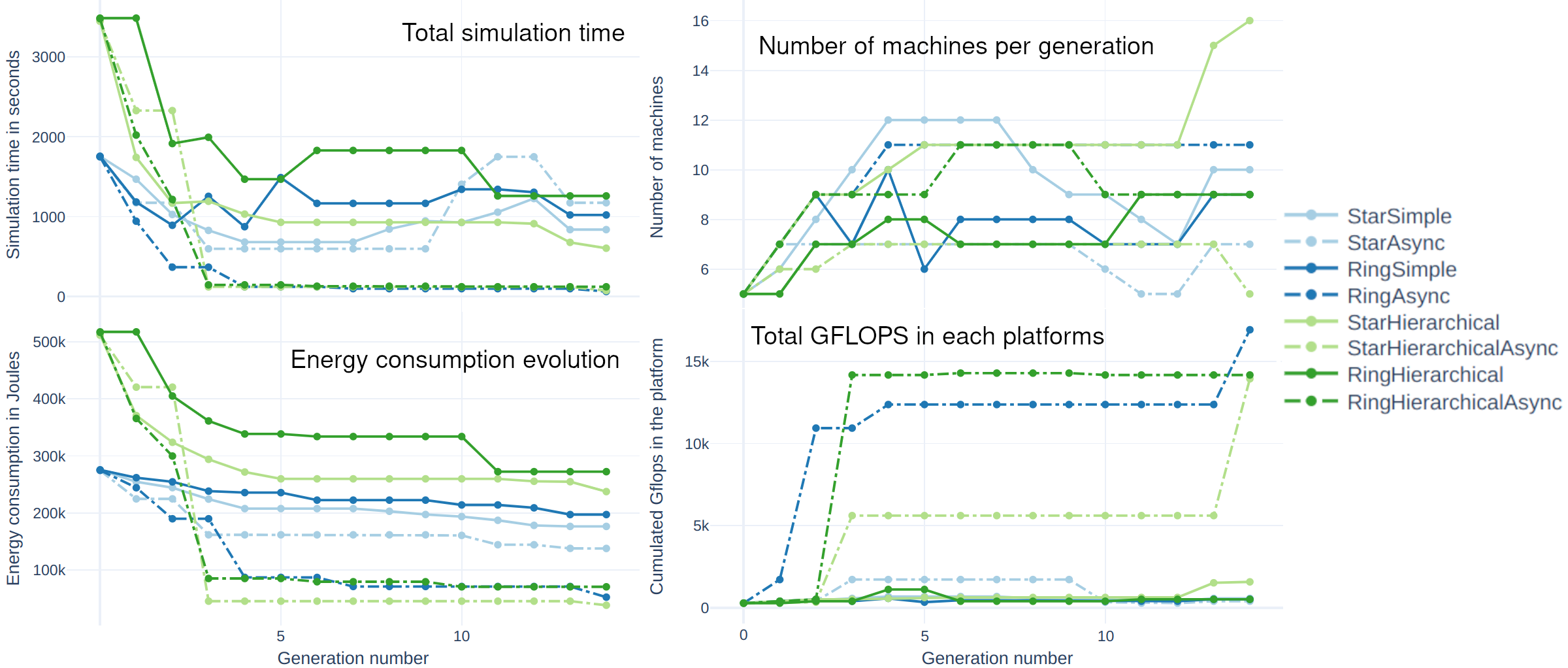}
    \caption{Results of the evolution algorithm performed using the total energy consumption as the optimization criteria. The plot represents the best individual of each configuration group per generation.}
    \label{fig:evolution-energy-consumption}
\end{figure}

\textbf{Results:} Fig.~\ref{fig:evolution-energy-consumption} shows the metrics of the best configuration in each group which represents a combination between a topology and a role algorithm. 
Here, the energy consumption is used as the optimization criteria, thus it can only decrease or keep steady. 
Several behaviours can be noticed, for example, increasing the number of machines can help lowering the consumption when the initial platform is under-scaled for the workload. 
Increasing the computational power of the platform (total \textit{GFLOPS}) gives the best results to save energy when using asynchronous algorithms. 
The heterogeneity of the platform can be tampered by asynchronism, e.g., the most powerful machines do not have to wait for the slowest ones, which reduces their idle time that have a non-negligible impact on consumption.
\section{Conclusion}
\label{sec:conclusion}
We have presented \texttt{Falafels}, a simulator capable of predicting the run-time and energy consumption of a~\ac{fl} system. 
The nearly instant execution differentiate our framework from the others and make it possible to apply optimization algorithms in order to choose the most frugal platform in terms of energy consumption.
However, the trade-off for fast executions is that the model convergence and accuracy cannot be known.
That is why we believe that this tool could be complementary to existing frameworks by providing platform related optimizations, in the same way grid search is used to find optimal hyper-parameters. 

As a future work, we aim to (a) extend the framework to other paradigms of~\ac{dai} and (b) support switching between discrete simulation and real execution. 
This last point will help us to verify and calibrate the simulator, but will also create opportunities to investigate the impact of platform optimization for energy reduction on model performance. 
Lastly, using the framework to study the impact of faults and the implementation of fault-tolerance algorithms on the energy consumption would be interesting.

\bibliographystyle{spmpsci}
\bibliography{references}

\end{document}